\documentstyle[11pt]{article}
\begin{document}
\thispagestyle{empty}
\begin{center}
\vspace{1.8cm} {\bf REPRESENTATIONS AND PROPERTIES OF GENERALIZED
$A_r$ STATISTICS, COHERENT STATES AND ROBERTSON UNCERTAINTY
RELATIONS}\\ \vspace{1.5cm} {\bf M.
Daoud}\footnote{m$_{-}$daoud@hotmail.com}\\ \vspace{0.5cm} {\it
Facult\'e des Sciences, D\'epartement de Physique, LPMC,\\
Agadir, Morocco}\\[1em]
\vspace{3cm} {\bf Abstract}
\end{center}
\baselineskip=18pt
\medskip
The generalization of $A_r$ statistics, including bosonic and
fermionic sectors, is performed by means of the so-called Jacobson
generators. The corresponding Fock spaces are constructed. The
Bargmann representations are also considered. For the bosonic
$A_r$ statistics, two inequivalent Bargmann realizations  are
developed. The first (resp. second) realization induces, in a
natural way, coherent states recognized as Gazeau-Klauder (resp.
Klauder-Perelomov) ones. In the fermionic case, the Bargamnn
realization leads to the Klauder-Perelomov coherent states. For
each considered realization, the inner product of two analytic
functions is defined in respect to a measure explicitly computed.
The Jacobson generators are realized as differential operators. It
is shown that the obtained coherent states minimize the
Robertson-Schr\"odinger uncertainty relation.

\newpage
\section{Introduction and motivations}
The quantum statistics, different from Bose and Fermi ones, have
attracted due attention in the literature [1-13] and various
versions were formulated. For example,  in two space dimensions,
one can have a one parameter family of statistics (anyons)
interpolating between bosons and fermions [4]. On the other hand,
in three and higher space dimensions the parastatistics, developed
by Green [1], constitute the natural extension of the usual Fermi
and Bose statistics. The interest on these exotic statistics is
mainly motivated by their promising applications in the theories
of fractional quantum Hall effect [7-8], anyonic
super-conductivity [9] and black hole statistics [10].  In the
Green generalization of conventional Bose and Fermi statistics,
the paraboson or parafermion algebra is generated by $r$ pairs of
creation and annihilation operators ($A_i^+$ , $A_i^-$) ($ i = 1,
2, \dots, r$) satisfying the trilinear relations (which replace
the standard bilinear commutation or anti-commutation relations)
$$ \big[[ A_i^+ , A_j^-]_{\pm}  , A_k^- ]\big]= - 2 \delta_{ik}
A_j^-,{\hskip 0.3cm}\big[[ A_i^+ , A_j^+]_{\pm}  , A_k^- ]\big]= - 2
\delta_{ik}
A_j^+ \mp 2 \delta_{jk}
A_i^+,{\hskip 0.3cm} \big[[ A_i^- , A_j^-]_{\pm}  , A_k^- ]\big]= 0
$$ \nonumber
where as usual $[x,y]_{\pm} = x y \pm y x$ and the
sign $+$ (resp. $-$) refer to parabosons (resp. parafermions).
It is interesting to mention that the para-Fermi relations are
associated with the orthogonal Lie algebra $so(2r+1) = B_r$ [14]
and the para-Bose statistics are connected to the orthosymplectic
superalgebra $osp(1/2r) = B(0,r)$ [15]. Recently, in view of this
connection between Lie algebras and super-algebras, a
classification of generalized quantum statistics were derived in
the framework of the classical Lie algebras $A_r$, $B_r$, $C_r$
and $D_r$ [6,16,17].\\ In this context, we shall be interested, in
the present paper, in generalized class of statistics associated
with the classical Lie algebra $A_r$. The general class of these
statistics is defined with the help of the notion of Lie triple
systems and the so-called Jacobson operators [18]. The latter
operators are known to be closely related to the description,
initiated by N. Jacobson, of Lie algebras by a minimal set of
generators and relations instead of to the well known Chevally
description. The second facet of this work concerns the Bargmann
representation associated with generalized $A_r$ statistics. The
latter is frequently important in the analysis of quantum field
theoretic systems and in connection with path integral methods.
Coherent states for $A_r$ statistics system emerges naturally in
the Bargmann realization. Coherent states for systems obeying
unconventional statistics were extensively investigated in
recent years. One may quote coherent states associated with
statistics developed in the context of quantum algebras like
$q$-bosons [19] and $k$-fermions [20]. Coherent states for
paraparticles were also constructed: Parabose coherent states were
proposed in [21] and Parafermi ones are given in [3,22]. All these
states appear to be quantum states closest to the classical ones.
The strongest qualitative measure of differences in the behavior
of quantum and classical properties is expressed by the Schr\"
odinger-Robertson uncertainty principle [23-24] (see also
[25-26]). As we are interested in the generalized $A_r$
statistics, it is natural to ask if the sets of coherent states,
which emerge in the construction of analytical representations,
minimize the Robertson-Shr\"odinger uncertainty relation. This
matter will constitute the last part of this work.\\ The paper is
organized as follows. Generalized quantum statistics is introduced
from a set of Jacobson generators (defined in section
2) satisfying certain triple relations. This generalization
includes two fundamental sectors. A fermionic one reproducing the
$A_r$ statistics introduced in [6]. The second sector is of
bosonic type. For each sector, we give the associated Fock space.
A Hamiltonian is derived in terms of the Jacobson generators
identified with creation and annihilation operators. In section 3,
the first analytic realization of the Fock space for the bosonic
$A_r$ statistics is performed. This realization generates the
so-called Gazeau-Klauder coherent states [27]. The second
realization, presented is section 4, leads to the
Klauder-Perelomov coherent states [28-29]. We also realize
analytically the Fock space related to the fermionic $A_r$
statistics. In this case, we show that the Jacobson generators act
on a over-complete set of coherent states similar to
Klauder-Perelomov ones labeling the complex projective spaces
${\bf CP}^r$. Differential actions of the Jacobson generators for
each obtained realization are given. In the last section, we show
that the quantum states, realizing analytically the vector states
of $A_r$ statistics, minimize the uncertainty principle. In other
words, they minimize the Robertson-Schr\"odinger uncertainty
relation. Some concluding remarks close this work.

\section{The generalized $A_r$ statistics}
In this section, we introduce the definitions of the Jacobson
operators and the generalized $A_r$ statistics viewed as Lie
triple system. We give the corresponding Fock space and a
Hamiltonian describing a quantum system obeying  generalized
$A_r$ statistics.
\subsection{Jacobson generators}
To begin, let us recall the definition of Lie triple systems. A
vector space with trilinear composition $[x,y,z]$ is called Lie
triple system if the following identities are satisfied:
$$[x,x,x] = 0,$$
$$[x,y,z] + [y,z,x] + [z,x,y] = 0,$$ $$[x,y,[u,v,w]] =
[[x,y,u],v,w] + [u,[x,y,v],w] + [u,v,[x,y,w]].$$\nonumber
According this definition, we will  introduce the generalized
$A_r$ statistics as Lie triple system. For this end, we consider
the set of $2r$ operators $x_i^+$ and $x_i^- $ ($i = 1, 2,..., r$) .
Inspired by the para-Fermi case [1] and
the example of $A_r$ statistics [6,16], these $2r$ operators
should satisfy certain conditions and relations. First, the
operators $x_i^+$ are mutually commuting. A similar statement
holds for the operators $x_i^-$. They also satisfy  the following
triple relations
\begin{equation}
\big[[ x_i^+ , x_j^- ] , x_k^+ ]\big] = - \epsilon  \delta_{jk}
x_i^+ - \epsilon  \delta_{ij} x_k^+
\end{equation}
\begin{equation}
\big[[ x_i^+ , x_j^- ] , x_k^- ]\big] =  \epsilon  \delta_{ik}
x_j^- + \epsilon  \delta_{ij}  x_k^-
\end{equation}
where $\epsilon \in {\bf R} \setminus \{0\}$. The algebra ${\cal A}$ (defined by
means of the generators $x_i^{\pm}$ and relations (1) and (2)) is
closed under the ternary operation $[x,y,z] = [[x,y],z]$ and
define a Lie triple system. Note that for $\epsilon = - 1$, the
algebra ${\cal A}$ reduces to one defining the $A_r$ statistics
discussed in [6]. The elements $x_i^{\pm}$ are referred to as
Jacobson generators which will be identified later with creation
and annihilation operators of a quantum system obeying
generalized $A_r$ statistics. We redefine the generators of the
algebra ${\cal A}$ as $a_i^{\pm} =
\frac{x_i^{\pm}}{\sqrt{|\epsilon|}}$. The triple relations (1) and
(2) may be rewritten as
\begin{equation}
\big[[ a_i^+ , a_j^- ] , a_k^+ ]\big] = - s  \delta_{jk}
a_i^+ - s  \delta_{ij} a_k^+,
\end{equation}
\begin{equation}
\big[[ a_i^+ , a_j^- ] , a_k^- ]\big] =  s  \delta_{ik}
a_j^- + s \delta_{ij}  a_k^-
\end{equation}
where $ s= \frac{\epsilon}{|\epsilon|}$ is the sign of the
parameter $\epsilon$ and $[ a_i^+ , a_j^+ ] = [ a_i^- , a_j^- ] =
0$. This redefinition is more convenient for our investigation, in
particular in determining the irreducible representation
associated with the algebra ${\cal A}$. As we will see in what
follows, the sign of the parameter $\epsilon$ play an importance in
the representation of the algebra ${\cal A}$ and consequently, one
can obtain different microscopic and macroscopic statistical
properties of the quantum system under consideration.
\subsection{The Hamiltonian}
To characterize a quantum gas obeying  the generalized $A_r$
statistics, we have to specify a Hamiltonian for the system. The
operators $a_i^{\pm}$ define creation and annihilation operators
for a quantum mechanical system, described by an Hamiltonian $H$,
when the Heisenberg equation of motion
\begin{equation}
[ H , a_i^{\pm} ] = \pm e_i a_i^{\pm}
\end{equation}
is  fulfilled. The quantities $e_i$ are the energies of the modes
$i = 1, 2, ... , r$. One can verify that if $|E \rangle$ is an
eigenstate with energy $E$, $a_i^{\pm}|E \rangle$ are eigenvectors
of $H$ with energies $E \pm e_i$. In this respect, the operators $
a_i^{\pm}$ can be interpreted as ones creating or annihilating
particles. To solve the consistency equation (5), we write the
Hamiltonian $H$ as
\begin{equation}
H = \sum_{i=1}^{r} e_i h_i
\end{equation}
which seems to be a simple sum of "free" (non-interacting)
Hamiltonians $h_i$. However, note that, in the quantum system
under consideration, the statistical interactions occur and are
encoded in the triple commutation relations (3) and (4).
Using the structure relations of the algebra ${\cal A}$, the
solution of the Heisenberg condition (5) is given by
\begin{equation}
h_i = \frac{s}{r+1}\bigg[ (r+1)[ a_i^- , a_i^+ ] - \sum_{j=1}^{r}[ a_j^- ,
a_j^+]
\bigg] + c
\end{equation}
where the constant $c$ will be defined later such that the
ground state (vacuum) of the Hamiltonian $H$ has zero energy.
\subsection{Fock representations}
We now consider a Hilbertian representation of the algebra ${\cal
A}$. Let  ${\cal F}$ be the Hilbert-Fock space on which the
generators of ${\cal A}$ act. Since, the algebra ${\cal A}$ is
spanned by $r$ pairs of Jacobson generators, it is natural to
assume that the Fock space is given by
\begin{equation}
{\cal F} =  \oplus_{n=0}^{\infty} {\cal H}^n,
\end{equation}
where ${\cal H}^n
\equiv \{ |n_1, n_2,\cdots , n_r\rangle\ , n_i \in {\bf{N}},
\sum_{i=1}^{r}n_i = n > 0\}$ and ${\cal H}^0 \equiv {\bf{C}}$.
The action of $a_i^{\pm}$, on ${\cal F}$, are defined by
\begin{equation}
a_i^{\pm} |n_1,\cdots, n_i,\cdots , n_r\rangle\ = \sqrt{F_i
(n_1,\cdots ,n_i \pm 1,\cdots, n_r)}|n_1,\cdots, n_i \pm 1,\cdots
, n_r\rangle\
\end{equation}
extended linearly, where the functions $F_i$ are called the structure functions
and are to be non-negatives so that all states are well defined.
To determine the expressions of the functions $F_i$ in terms of
the quantum numbers $n_1, n_2, \cdots, n_r$, let first assume that
 $a_i^- |0, 0,\cdots , 0\rangle\ = 0$
for all $i = 1, 2, \cdots, r$. This implies that the functions
$F_i$ satisfy
\begin{equation}
F_i (n_1,\cdots ,n_i,\cdots, n_r) = n_i  G_i (n_1,\cdots
,n_i,\cdots, n_r),
\end{equation}
in a factorized form where the new functions $G_i$ are
defined such that $G_i (n_1,\cdots ,n_i = 0,\cdots,
n_r)\neq 0$ for $i=1, 2,\cdots, r$. Furthermore, since the
Jacobson operators satisfy the trilinear relations (3) and (4),
these functions should be affine in the quantum numbers $n_i$:
\begin{equation}
G_i (n_1,\cdots ,n_i,\cdots, n_r)= k_0 + (k_1 n_1 + k_2
n_2 + \cdots + k_r n_r ).
\end{equation}
Finally, using the relations $[a_i^+ , a_j^+] = 0$ and $\big[[
a_i^+ , a_i^- ] , a_i^+ ]\big] = - 2s a_i^+ $, one obtain $k_i =
k_j $ and $k_i = s$, respectively. For convenience, we set  $k_0 =
k - \frac{1+s}{2}$ assumed to be a non-vanishing integer. The
actions of the Jacobson generators on the
states spanning the Hilbert-Fock space ${\cal F}$ are now given by
\begin{equation}
a_i^{-} |n_1,\cdots, n_i,\cdots , n_r\rangle\ = \sqrt{n_i (k_0 +
s(n_1+n_2+\cdots+n_r))}|n_1,\cdots, n_i-1,\cdots , n_r\rangle,
\end{equation}
\begin{equation}
a_i^{+} |n_1,\cdots, n_i,\cdots , n_r\rangle\ = \sqrt{(n_i+1) (k_0
+ s(n_1+n_2+\cdots+n_r+1))}|n_1,\cdots, n_i+1,\cdots , n_r\rangle.
\end{equation}
The dimension of the irreducible representation space ${\cal F}$
is determined by the  condition:
\begin{equation}
k_0 + s(n_1+n_2+\cdots+n_r) > 0.
\end{equation}
It depends on the sign of the parameter $s$. It is clear that for
$s=1$, the Fock space ${\cal F}$ is infinite dimensional. However,
for $s=-1$, there exists a finite number of basis states satisfying the
condition $n_1+n_2+\cdots+n_r \leq k-1$. The dimension is given,
in this case, by $\frac{(k-1+r)!}{(k-1)!r!}$. This is exactly the
dimension of the Fock representation of $A_r$ statistics discussed
in [6]. This condition-restriction is closely related to so-called
generalized exclusion Pauli principle according to which no more
than $k-1$ particles can be accommodated in the same quantum
state. In this sense, for $ s = -1$, the generalized $A_r$ quantum
statistics give statistics of fermionic behaviour. They will be
termed here as fermionic $A_r$ statistics and ones corresponding
to $s = 1$ will be named bosonic $A_r$ statistics.\\
Setting $c =
\frac{r}{r+1} s k_0$ in (7) and using the equation (6) together
with the actions of creation and annihilation operators (12-13),
one has
\begin{equation}
H |n_1,\cdots, n_i,\cdots , n_r\rangle\ = \sum_{i=1}^{r}e_i
n_i|n_1,\cdots, n_i,\cdots , n_r\rangle.
\end{equation}
It is remarkable that, for $s=-1$, the spectrum of $H$ is similar
(with a slight modification) to energy eigenvalues of the $A_r$
Calogero model (see for instance Eq.(1.2) in [30]). The latter
describe the dynamical model containing $r+1$ particles on a line
with long rang interactions and provides a microscopic realization
of fractional statistics [13,31].\\ Finally, we point out one
interesting property of the generalized $A_r$ statistics.
Introduce the operators $b_i^{\pm} = \frac{a_i^{\pm}}{\sqrt{k}}$
for $i = 1, 2, \cdots, r$ and consider $k$ very large. From
equations (12) and (13), we obtain
\begin{equation}
b_i^{-} |n_1,\cdots, n_i,\cdots , n_r\rangle\ \approx
\sqrt{n_i}|n_1,\cdots, n_i-1,\cdots , n_r\rangle,
\end{equation}
\begin{equation}
b_i^{+} |n_1,\cdots, n_i,\cdots , n_r\rangle\ \approx
\sqrt{n_i+1}|n_1,\cdots, n_i+1,\cdots , n_r\rangle.
\end{equation}
In this limit, the generalized $A_r$ statistics (fermionic and
bosonic ones) coincide with th Bose statistics and the Jacobson
operators reduce to Bose ones (creation and annihilation operators
of harmonic oscillators).\\
Besides the Fock representation
discussed in this section, it is interesting to look for
analytical realizations of the space representation associated
with the Fock representations of the generalized $A_r$ statistics.
These realizations  constitute an useful analytical tool in
connection with variational and path integral methods to describe
the quantum dynamics of the system described by the Hamiltonian
$H$.

\section{Bargmann realization and Gazeau-Klauder coherent states}
This section is devoted to a realization \`a la Bargmann using a
suitably defined Hilbert space of entire analytic functions
associated with the bosonic $A_r$ statistics introduced above. In
this first analytic realization, the Jacobson creation operators
are realized as simple multiplication by some complex variables.
As by product, this realization generates, in a natural way, the
Gazeau-Klauder coherent states associated to a quantum mechanical
system described by the Hamiltonian given by (6) and (7) for the
particular case $s = 1$. To begin, we realize the vectors $|k;
n_1,\cdots,n_r\rangle$ as powers of complex variables
$\omega_1,\cdots,\omega_r$ on which the Jacobson creations
operators $a_i^+$ act as multiplication by $\omega_i$
\begin{equation}
|k; n_1,\cdots,n_r\rangle \longrightarrow C_{k; n_1,\cdots,n_r}
\omega_1^{n_1}\cdots\omega_r^{n_r}
\end{equation}
where the set of coefficients $C_{k; n_1,\cdots,n_r}$ occurring in
the last expression will be determined in what follows. Equation
(13) leads to the following recursion relation
\begin{equation}
C_{k; n_1,\cdots,n_i,\cdots,n_r} =
((n_i+1)(k+n_1+\cdots+n_i+\cdots+n_r))^{\frac{1}{2}}
C_{k;n_1,\cdots,n_i+1,\cdots,n_r}.
\end{equation}
Solving this equation, we obtain
\begin{equation}
C_{k; n_1,\cdots,n_i,\cdots,n_r} = \bigg[{\frac{(k-1 + n - n_i)!
}{n_i!(k-1 + n) !}}\bigg]^{\frac{1}{2}} C_{k;
n_1,\cdots,0,\cdots,n_r}.
\end{equation}
where $n = n_1 + n_2 + \cdots + n_r$. We repeat this procedure for
all $i = 1, 2,\cdots,r$ and setting $C_{k; 0,\cdots,0} = 1$, we
obtain
\begin{equation}
C_{k; n_1,\cdots,n_i,\cdots,n_r} = \bigg[{\frac{(k-1)!
}{n_1!\cdots n_r!(k-1 + n) !}}\bigg]^{\frac{1}{2}}.
\end{equation}
If we define the operators $N_i$ ($\neq a_i^+ a_i^-$) such that
\begin{equation}
N_i|k;n_1,\cdots,n_i\cdots,n_r\rangle = n_i
|k;n_1,\cdots,n_i\cdots,n_r\rangle,
\end{equation}
it is easy to see that the operators $N_i$ act in this
differential realization as
\begin{equation}
N_i \longrightarrow \omega_i \frac{\partial}{\partial\omega_i}.
\end{equation}
To define the differential actions of the annihilation operators
$a_i^-$, we use their actions on the Fock space (Eq.12) together
with the equation (23). One has
\begin{equation}
a_i^- \longrightarrow k \frac{\partial}{\partial\omega_i} +
\omega_i
\frac{\partial^2}{\partial^2\omega_i} +
\frac{\partial}{\partial\omega_i}\sum_{i \neq
j}\omega_j \frac{\partial}{\partial\omega_j}.
\end{equation}
A general vector
\begin{equation}
|\psi\rangle = \sum_{n_1,\cdots,n_r}
\psi_{n_1,\cdots,n_r}|k;n_1,\cdots,n_r\rangle
\end{equation}
in the Fock space ${\cal F}$ now is realized as follows
\begin{equation}
\psi(\omega_1,\cdots,\omega_r) = \sum_{n_1,\cdots,n_r}
\psi_{n_1,\cdots,n_r}C_{k;n_1,\cdots,n_r}\omega_1^{n_1}\cdots\omega_r^{n_r},
\end{equation}
a.e. We define the inner product in this realization in the following
form
\begin{equation}
\langle\psi'|\psi\rangle = \int d^2\omega_1 \cdots d^2\omega_r
K(k;\omega_1,\cdots,\omega_r)\psi'^{\star}(\omega_1,\cdots,\omega_r)
\psi(\omega_1,\cdots,\omega_r)
\end{equation}
where $d^2\omega_i \equiv dRe\omega_idIm\omega_i$,
where $K$ is to be determined and the
integration extends over the entire complex space $\bf{C}^r$. To
compute the density function $K$, appearing in the definition of the
inner product (27), we choose  $|\psi\rangle$(resp.
$|\psi'\rangle$) to be the vector $|k;n_1,\cdots,n_r\rangle$
(resp. $|k;n'_1,\cdots,n'_r\rangle$). We also assume that $K$
depends only on $\rho_i = |\omega_i|$ for $i = 1,\cdots, r$. This
assumption reflects the isotropic condition used in the moment
problems. Then, it is a simple matter of computation to show that the
function $K(k;\rho_1,\cdots,\rho_r)$ should satisfy the integral
equation
\begin{equation}
(2\pi)^r\int_0^{\infty}\cdots\int_0^{\infty} d\rho_1\cdots d\rho_r
K(k;\rho_1,\cdots ,\rho_r) |\rho_1|^{2n_1+1}\cdots
|\rho_r|^{2n_r+1}= \frac{n_1!\cdots n_r!(k-1+n)!}{(k-1)!}.
\end{equation}
A solution of this equation exists [32] (see a nice proof in [33])
in term of the Bessel function
\begin{equation}
K(k;R) = \frac{2}{\pi^r(k-1)!}R^{k-r}K_{k-r}(2R)
\end{equation}
where $R^2 = \rho_1^2 + \cdots + \rho_r^2$. Note that the analytic
function $\psi (\omega_1, \cdots ,\omega_r)$ can be viewed as the
inner product of the ket $|\psi\rangle$ with a bra $\langle k;
\omega_1^{\star}, \cdots ,\omega_r^{\star}|$ labeled by the
complex conjugate of the variables $\omega_1, \cdots ,\omega_r$
\begin{equation}
\psi (\omega_1, \cdots ,\omega_r) = {\cal N}\langle k;
\omega_1^{\star}, \cdots ,\omega_r^{\star}|\psi \rangle
\end{equation}
where ${\cal N} \equiv {\cal N}(|\omega_1|, \cdots ,|\omega_r|)$
stands for a normalization constant of the states $|k; \omega_1,
\cdots ,\omega_r\rangle$ to be adjusted later. As a particular case,
if we take $|\psi \rangle = |k; n_1,\cdots,n_r\rangle$, we get
\begin{equation}
\langle k; \omega_1^{\star}, \cdots ,\omega_r^{\star}|k;
n_1,\cdots,n_r \rangle = {\cal N}^{-1} C_{k;n_1,\cdots,n_r}
\omega_1^{n_1} \cdots \omega_r^{n_r}.
\end{equation}
The last equation implies
\begin{equation}
|k; \omega_1, \cdots ,\omega_r\rangle = {\cal N}^{-1}
\sum_{n_1=0}^{\infty} \cdots \sum_{n_r=0}^{\infty}
\bigg[{\frac{(k-1)! }{n_1!\cdots n_r!(k-1 + n)
!}}\bigg]^{\frac{1}{2}} \omega_1^{n_1} \cdots \omega_r^{n_r}
\end{equation}
where the normalization constant ${\cal N}$ is
\begin{equation}
{\cal N}^2(|\omega_1|, \cdots ,|\omega_r|) = \sum_{n_1=0}^{\infty}
\cdots \sum_{n_r=0}^{\infty} {\frac{(k-1)! }{n_1!\cdots n_r!(k-1 +
n) !}} |\omega_1|^{2n_1} \cdots |\omega_r|^{2n_r}
\end{equation}
The states $|k; \omega_1, \cdots ,\omega_r\rangle$ are not
orthogonal and constitute an over-complete set with respect to the
measure given by (29). It is also interesting to remark that they
are eigenvectors of the Jacobson operators $a_i^-$ with the
eigenvalue $\omega_i$. In this sense, the states $|k; \omega_1,
\cdots ,\omega_r\rangle$ can be considered as Gazeau-Klauder
coherent states associated with  a quantum mechanical system whose
Hamiltonian is given by (6) and (7).
\section{Bargmann realization and Klauder-Perelomov coherent states}
\subsection{Bosonic $A_r$ statistics}
Here, we shall consider the second analytic realization associated
with bosonic $A_r$ statistics. We consider the complex domain
${\cal D} = \{(z_1, z_2, \cdots, z_r ): |z_1|^2 + |z_2|^2 + \cdots
+ |z_r|^2 < 1 \}$ . The reason for this condition will be clarified
in the sequel of this subsection. In this realization, the
annihilation operators $a_i^-$ are represented as derivation with
respect to the complex variables $z_i$
\begin{equation}
a_i^-\longrightarrow \frac{\partial}{\partial z_i}
\end{equation}
and the basis elements of the Fock space are realized as follows
\begin{equation}
|k; n_1, \cdots , n_r\rangle  \longrightarrow C_{k; n_1,\cdots
,n_r} z_1^{n_1} \cdots z_r^{n_r}.
\end{equation}
Using the action of the annihilation operators on the Fock space
${\cal F}$ and the correspondence (35), one obtain the following
recursion formula
\begin{equation}
\sqrt{k-1+n_1+\cdots +n_i+ \cdots +n_r}C_{k; n_1,\cdots,n_i-1,
\cdots,n_r}= \sqrt{n_i}C_{k; n_1,\cdots,n_i,
\cdots,n_r}
\end{equation}
which can be solved in a similar manner that one used above
(Eq.19) and setting $C_{k; 0,\cdots, 0} = 1$. We have
\begin{equation}
C_{k; n_1,\cdots,n_i,
\cdots,n_r}= \bigg[{\frac{(k-1 + n)!
}{n_1!\cdots n_r! (k-1)!}}\bigg]^{\frac{1}{2}}
\end{equation}
where $n = n_1 + n_2 + \cdots + n_r$. Having the expression of the
coefficients $C$, one can determine the differential action of the
Jacobson creation operators. Indeed, using the actions of the
generators $a_i^+$ on the Fock space and the triple relations (3)
and (4) , we show that
\begin{equation}
a_i^+ \longrightarrow k z_i + z_i \sum_{j=1}^r z_j\frac{\partial}{\partial z_j};
\end{equation}
i.e., the Jacobson generators act as first order linear differential
operators. Here also, we realize a general vector of the Fock
space ${\cal F}$ ($s=1$) $$|\phi \rangle =
\sum_{n_1=0}^{\infty}\sum_{n_2=0}^{\infty}\cdots
\sum_{n_r=0}^{\infty} \phi_{n_1, n_2 \cdots , n_r}|k; n_1,n_2,
\cdots ,n_r \rangle$$\nonumber\\ as
\begin{equation}
\phi(z_1, z_2 \cdots , z_r) =
\sum_{n_1=0}^{\infty}\sum_{n_2=0}^{\infty}\cdots
\sum_{n_r=0}^{\infty} \phi_{n_1, n_2 \cdots , n_r} C_{k;n_1, n_2
\cdots , n_r} z_1^{n_1} z_2^{n_2} \cdots z_r^{n_r},
\end{equation}
a.e. The inner product of two functions $\phi$ and $\phi'$ is defined
now as follows
\begin{equation}
\langle\phi'|\phi\rangle = \int \int \cdots \int d^2z_1  d^2z_2
\cdots d^2z_r \Sigma(k;z_1,z_2 \cdots ,z_r)\phi'^{\star}(z_1,z_2
\cdots ,z_r) \phi(z_1,z_2 \cdots ,z_r)
\end{equation}
where the integration is carried out the complex domain ${\cal
D}$. The computation of the integration measure $\Sigma$, assumed
to be isotropic , can be performed by choosing $|\phi\rangle = |k;
n_1,n_2, \cdots ,n_r \rangle $ and $|\phi'\rangle = |k; n'_1,n'_2,
\cdots ,n'_r \rangle $. It follows that the measure $\Sigma$
satisfy the following moment equation
\begin{equation}
\int \int \cdots \int d\varrho_1  d\varrho_2
\cdots d\varrho_r \Sigma(k;\varrho_1,\varrho_2, \cdots ,\varrho_r)
\varrho_1^{2n_1+1}\varrho_2^{2n_2+1} \cdots \varrho_r^{2n_r+1} =
\frac{n_1!n_2!\cdots n_r! (k-1)!}{(2\pi)^r(k-1+n)!}
\end{equation}
where $n = n_1+n_2+\cdots +n_r$ and $\varrho_i = |z_i|$. To find
the isotropic function satisfying the equation (41), we use following result
$$\int_0^1 t_1^{n_1}dt_1  \int_0^{1-t_1}
t_2^{n_2}dt_2 \cdots  \int_0^{1-t_1-t_2-\cdots
-t_{r-1}}
t_r^{n_r}(1-t_1-t_2-\cdots
-t_r)^{k-r-1}dt_r$$ \nonumber
\begin{equation}
= \frac{n_1!n_2!\cdots n_r!(k-1)!}{(k-1+n)! (k-r)(k-r+1)\cdots
(k-1)}
\end{equation}
which can be easily verified. The measure is then given by
\begin{equation}
\Sigma (k;\varrho_1,\varrho_2, \cdots ,\varrho_r) =
\pi^{-r}(k-r)(k-r+1)\cdots (k-1)[1 - (\varrho_1^2 + \varrho_2^2+
\cdots + \varrho_r^2)]^{k-r-1}.
\end{equation}
One can write the function $\phi(z_1, z_2 \cdots , z_r)$ as the
product of the state $|\phi\rangle $ with some ket $|k; z^*_1,
z^*_2 \cdots , z^*_r \rangle$ labeled by the complex conjugate of
the variables $z_1, z_2, \cdots , z_r$
\begin{equation}
\phi(z_1, z_2, \cdots , z_r)= {\cal N} \langle k; z^*_1, z^*_2,
\cdots , z^*_r |\phi \rangle.
\end{equation}
Taking $|\phi\rangle = |k; n_1, n_2, \cdots , n_r \rangle$, we
have
\begin{equation}
|k; z_1, z_2, \cdots , z_r \rangle =  {\cal N}^{-1}
\sum_{n_1=0}^{\infty}\sum_{n_2=0}^{\infty}\cdots
\sum_{n_r=0}^{\infty} \bigg[{\frac{(k-1 + n)! }{n_1!\cdots n_r!
(k-1)!}}\bigg]^{\frac{1}{2}} z_1^{n_1} z_2^{n_2} \cdots z_r^{n_r}.
\end{equation}
The expansion (45) converges when $|z_1|^2+ |z_2|^2+ \cdots +
|z_r|^2 < 1$. In other words, the complex variables $z_1, z_2,
\cdots ,z_r $ should be in the complex domain ${\cal D}$ defined
above. The normalization constant in (45) is given by
\begin{equation}
{\cal N} = (1 - |z_1|^2- |z_2|^2- \cdots - |z_r|^2)^\frac{k}{2}.
\end{equation}
The states (45) are continuous in the labeling, constitute an over
complete set in the respect to the measure given by (43) and then
are coherent in the Klauder-Perelomov sense. It comes that the
quantum states of bosonic  $A_r$ statistics system admit two
non-equivalent realizations.

\subsection{Fermionic $A_r$ statistics}
Now, we construct the analytic realization of the irreducible
representation related to fermionic $A_r$ statistics $(s = -1)$
characterized by the so-called  generalized Pauli principle.
First, note that since the Fock space is of finite dimension, the
Jacobson creation operators can not be represented as a
multiplication by some complex variable. Unlike the bosonic  $A_r$
statistics, only one realization can be made in this case. It
corresponds to one in which the generators $a_i^-$ act as
\begin{equation}
a_i^- \longrightarrow \frac{\partial}{\partial\zeta_i}
\end{equation}
in the space of the polynomials of the form $C_{k; n_1, n_2,\cdots
, n_r } \zeta_1^{n_1}\zeta_2^{n_2} \cdots \zeta_r^{n_r}$ in the
$r$-dimensional space $\bf{C}^r$of complex lines $(\zeta_1,
\zeta_2, \cdots , \zeta_r)$ with
\begin{equation}
|k; n_1, n_2,\cdots , n_r  \rangle\longrightarrow C_{k; n_1,
n_2,\cdots , n_r } \zeta_1^{n_1}\zeta_2^{n_2} \cdots \zeta_r^{n_r},
\end{equation}
a.e. The coefficients in (48) satisfy the recurrence formula
\begin{equation}
\sqrt{n_i}C_{k; n_1, n_2,\cdots ,n_i,\cdots ,n_r } =\sqrt{k-n}
C_{k; n_1, n_2,\cdots ,n_i-1,\cdots , n_r }
\end{equation}
where $n = n_1 + n_2 + \cdots + n_r$. The solution, for all $i=1,
2, \cdots , r$, is given by
\begin{equation}
C_{k; n_1, n_2,\cdots ,n_r } = \bigg[ \frac{(k-1)!}{n_1!n_2!\cdots
n_r!(k-1-r)} \bigg]^{\frac{1}{2}}.
\end{equation}
The creation generators $a_i^+$ act in this realization as
\begin{equation}
a_i^- \longrightarrow (k-1)\zeta_i - \zeta_i \sum_{j=1}^{r}\zeta_j
\frac{\partial}{\partial\zeta_i};
\end{equation}
i.e; first order differential operators. \\
As in the previous cases,
there exists a measure $\sigma (k; \zeta_1, \zeta_2, \cdots ,
\zeta_r)$ by means of which one can define the inner product
between two arbitrary functions. To compute this measure, we use
the orthogonality of the Fock states $|k; n_1, n_2, \cdots ,n_r
\rangle$ which gives $$\int \int \cdots \int d^2\zeta_1 d^2\zeta_2
\cdots d^2\zeta_r \sigma (k; \zeta_1, \zeta_2, \cdots , \zeta_r)
C_{k; n_1, n_2, \cdots ,n_r}C_{k; n'_1, n'_2, \cdots ,n'_r}
\zeta_1^{n_1}\zeta_1^{n'_1} \zeta_2^{n_2}\zeta_2^{n'_2} \cdots
\zeta_r^{n_r}\zeta_r^{n'_r}$$\nonumber
\begin{equation}
= \delta_{n_1,n'_1}\delta_{n_2,n'_2}
\cdots  \delta_{n_r,n'_r}
\end{equation}
Setting $\zeta_i = |\zeta_i|e^{i\theta}$ and assuming the isotropy
of the measure, the relation (52) becomes $$ \int_0^{\infty}
\int_0^{\infty} \cdots \int_0^{\infty}dx_1 dx_2 \cdots dx_r\mu(k,
x_1 ,x_2, \cdots ,x_r) x_1^{n_1}x_2^{n_2}\cdots x_r^{n_r} $$
\begin{equation}
= \frac{n_1!n_2!\cdots n_r!(k-1-n)!}{(k-1)!}
\end{equation}
where $\mu \equiv \pi^r \sigma$ and $x_i = |\zeta_i|^2$. Using the
Mellin inverse transform [32], one obtain
\begin{equation}
\mu(k, x_1 ,x_2, \cdots ,x_r) = \frac{(k-1+r)!}{(k-1)!}(1 + x_1 +
x_2 + \cdots + x_r )^{-(k+r)}.
\end{equation}
Any function $f(\zeta_1, \zeta_2, \cdots , \zeta_r)$ can be
written in the following form
\begin{equation}
f(\zeta_1, \zeta_2, \cdots , \zeta_r) = {\cal N} \langle k;
\zeta_1^*, \zeta_2^*, \cdots , \zeta_r^* | f \rangle
\end{equation}
where $| f \rangle$ is a generic element of the Fock space and the
normalization constant is given by
\begin{equation}
{\cal N}(|\zeta_1|^2 , |\zeta_2|^2 , \cdots , |\zeta_r|^2) = (1 +
|\zeta_1|^2 + |\zeta_2|^2 + \cdots + |\zeta_r|^2
)^{-\frac{k-1}{2}}.
\end{equation}
It is interesting to note that the states $|k;\zeta_1, \zeta_2,
\cdots , \zeta_r \rangle$ are nothing but the coherent states
parameterizing the complex projective space $\mathbf{CP}^r$. They
were used in the description of quantum Hall systems in higher
dimension complex projective spaces [34]. In this respect, we
believe that the generalized quantum $A_r$ statistics can be
linked to this subject.
\section{Robertson-Schr\"odinger uncertainty relation}
The main aim of this section is to show that the coherent states,
derived in the previous section, minimize the
Robertson-Schr\"odinger uncertainty relation [23-24]. The states
minimizing this relation are called
minimum uncertainty states (or
intelligent states) [25-26].
For this end, we recall that for $2r$ observables (hermitian
operators) $(X_1,X_2, \cdots , X_{2r} ) \equiv X$, Robertson
established the following uncertainty relation for the matrix
dispersion $\sigma$
\begin{equation}
\det \sigma(X) \geq \det C(X)
\end{equation}
where $\sigma_{\alpha \beta} = \frac{1}{2}\langle
X_{\alpha}X_{\beta} + X_{\beta}X_{\alpha} \rangle - \langle
X_{\alpha}X_{\beta} \rangle$, $(\alpha = 1, 2, \cdots , 2r )$, and
$C$ is the antisymmetric matrix of the mean commutators $C_{\alpha
\beta} = -\frac{i}{2}[X_{\alpha} , X_{\beta}]$. Here $\langle O
\rangle$ stands for the mean value of the operator $O$ in a given
quantum state which is generally a mixed state. For $r=1$,
inequality (57) coincides with Schr\"odinger uncertainty relation
which gives the Heisenberg uncertainty relation when the
term $\sigma_{12}$ is vanishing.
\subsection{Gazeau-Klauder coherent states}
To show that the Gazeau-Klauder coherent states (32) minimize the
uncertainty relation (57),i.e. $\det \sigma(X) =  \det C(X)$, let
define the hermitian operators $(X_1,X_2, \cdots , X_{2r})$ as
\begin{equation}
X_i = \frac{1}{2} (a_i^+ + a_i^-) {\hskip 1cm} X_{i+r} =
\frac{i}{2} (a_i^+ - a_i^-)
\end{equation}
in terms of the creation and annihilation operators of the quantum
system described by the Hamiltonian $H$ .\\ The matrix $A \equiv
(a_1^-, a_2^-, \cdots, a_r^-, a_1^+, a_2^+, \cdots, a_r^+)$ is
related to $X$ as $X = U A$
\begin{displaymath}
U = \frac{1}{2}\left( \begin{array}{ccc} {\bf 1}_r & {\bf 1}_r \\
-i {\bf 1}_r & i {\bf 1}_r
\\
\end{array} \right)
\end{displaymath}
where ${\bf 1}_r$ is $r\times r$ unit matrix. It follows that both
matrices  $\sigma (X)$ and $C(X)$  can be expressed in terms of
matrices  $\sigma (A)$ and $C(A)$:
\begin{equation}
\sigma (X) = U \sigma (A) U^T {\hskip 1cm} C(X) = U C(A) U^T
\end{equation}
The eigenvalue equations $a_i^- |\omega_1, \omega_2, \cdots,
\omega_r \rangle = \omega_i |\omega_1, \omega_2,\cdots, \omega_r
\rangle$, provide us with the following relations between the
matrix elements of $\sigma (A)$ and $C(A)$
\begin{equation}
\sigma_{ij} = 0 {\hskip 1cm} C_{ij} = 0
\end{equation}
\begin{equation}
\sigma_{i+r,j+r} = 0 {\hskip 1cm} C_{i+r,j+r} = 0
\end{equation}
\begin{equation}
\sigma_{i,j+r} =  i C_{i,j+r} {\hskip 1cm} \sigma_{i+r,j} = -i
C_{i+r,j}.
\end{equation}
The last relations give $\det \sigma(A) =  \det C(A)$ which in
view of (59) (and the nondegeneracy of $U$, $\det U =
{(\frac{i}{2})}^r $ ) leads to the needed equality in the
Robertson-Schr\"odinger uncertainty relation (57), namely $\det
\sigma(X) = \det C(X)$.
\subsection{Klauder-Perelomov coherent states}
Now, it remains to show that the Klauder-Perelomov, for bosonic
and fermionic $A_r$ statistics, minimize
the Robertson-Schr\"odinger uncertainty relation. We first write
the coherent states (45) and (55) as resulting from the action of
some displacement operator on the lowest weight state (the
vacuum). Indeed, by a more or less complicated calculus, one can
show that coherent states (45) coincide with the vectors
\begin{equation}
{\cal D}(\eta_1, \eta_2,\cdots ,\eta_r) |0, 0, \cdots ,0 \rangle =
{\cal D}(\eta_r)\cdots {\cal D}(\eta_2) {\cal D}(\eta_1)|0, 0,
\cdots ,0\rangle\
\end{equation}
where the displacement operators ${\cal D}(\eta_i)$ are defined by
\begin{equation}
{\cal D}(\eta_1) = \exp (\eta_1 a_1^+ - \overline{\eta}_1a_1^-)
{\hskip 0.5cm}, {\hskip 0.5cm}{\cal D}(\eta_i) = \exp (\eta_i [a_{i-1}^- , a_{i}^+]
- \overline{\eta}_i [ a_{i}^- , a_{i-1}^+ ])
\end{equation}
for $i = 2, 3, \cdots, r $. The complex parameters occurring in
the equations (63) and (64) are given in terms of variables
labeling the coherent states (45) as  $\tanh^2|\eta_1| =
|z_1|^2+|z_2|^2+ \cdots + |z_r|^2$ , $\tan^2|\eta_i| =
|z_{i-1}|^{-2} (|z_i|^2+|z_{i+1}|^2+ \cdots + |z_r|^2 )$ for $i =
2, 3, \cdots, r$ and $\frac{\eta_i}{|\eta_i|} =
\frac{z_i}{|z_i|}$.\\ Similarly, the coherent states obtained for
fermionic $A_r$ statistics derived from the expression (55) can be
written as
\begin{equation}
{\cal D}(\eta'_1, \eta'_2,\cdots ,\eta'_r) |0, 0, \cdots ,0
\rangle = {\cal D}(\eta'_r)\cdots {\cal D}(\eta'_2) {\cal
D}(\eta'_1)|0, 0, \cdots ,0\rangle\
\end{equation}
where the unitary operators ${\cal D}(\eta'_i)$ are defined as
follows
\begin{equation}
{\cal D}(\eta'_1) = \exp (\eta'_1 a_1^+ - \overline{\eta'}_1a_1^-)
{\hskip 0.5cm}, {\hskip 0.5cm} {\cal D}(\eta'_i) = \exp (\eta'_i [a_{i}^+ ,
a_{i-1}^-] - \overline{\eta'}_i [ a_{i-1}^+ , a_{i}^- ]).
\end{equation}
The complex variables $\zeta_i$, labeling the fermionic $A_r$
statistics states, are related to ones, parameterizing the
displacement operators (66), as follows $\zeta_i = Z_1Z_2 \cdots
Z_i$, ($i=1,2, \cdots, r$) with $Z_j =
\frac{\eta_j'}{|\eta_j'|}\tan |\eta_j'|\cos|\eta_{j+1}'|$ for
$j=1, 2, \cdots , r-1$ and $Z_r = \frac{\eta_r'}{|\eta_r'|}\tan
|\eta_r'|$.\\ To prove that the states (63) and (65) minimize the
Robertson-Schr\"odinger uncertainty relation, we  shall show that
they are  eigenstates of the linear combination of Jacobson
generators $A_i^- \equiv A_i^- (u,v)= u_{ij}a_j^- + v_{ij}a_j^+$ (summation over
repeated indices). To simplify our notations, we denote by
$|coh,s=\pm1\rangle$  the coherent states for bosonic ($s=1$) and
fermionic ($s=-1$) $A_r$ statistics. Using the triple relation
commutation, one get
\begin{equation}
{\cal D}^{\dagger} a_i^+ {\cal D} = x_{ij} a_j^-
+ y_{ij} a_j^+ + z_{ijk} [a_j^- , a_k^+]
\end{equation}
where ${\cal D}$ is given by (64) (resp. (66)) for bosonic  $A_r$
statistics (resp. fermionic $A_r$ statistics ). The complex
parameters $x_{ij}$, $y_{ij}$ and $z_{ijk}$ are functions of the
variables labeling the coherent states (The expressions of $x_{ij}$,
$y_{ij}$ and $z_{ijk}$ can be obtained by using the trilinear relations
(3) and (4) coupled with Baker-Campbell-Hausdorff relation). From (67), one obtain
$${\cal D}^{\dagger}A_i^-{\cal D}|0, 0, \dots , 0\rangle
=$$\nonumber
$$\bigg[
(u_{ij} x_{jk} + v_{ij}y_{jk}^*) a_k^-
+ (u_{ij} y_{jk} + v_{ij}x_{jk}^*) a_j^+
+ (u_{ij} z_{jkl} + v_{ij}z_{jkl}^*) [a_k^- , a_l^+]\bigg]
|0, 0, \dots , 0\rangle$$
Since $a_j^-|0, 0, \dots , 0\rangle = 0$ and
$[a_k^- , a_l^+]|0, 0, \dots , 0\rangle = (k+ \frac{s-1}{2}) \delta_{kl}|0,
0, \dots ,
0\rangle$, the coherent states (63) and (65) are
eigenstates of $A_i^-$ if the $r \times r$ matrices $u, v, x $ and
$y$ satisfy the condition $u y + v x^* = 0$ and we have
\begin{equation}
A_i^- |coh, s= \pm 1\rangle = (k+ \frac{s-1}{2}) \sum_{jl} (u_{ij}
z_{jll} + v_{ij}z_{jll}^*)|coh, s= \pm 1\rangle
\end{equation}
In the last
step in our proof, we consider the quadrature components $X_i$ and
$X_{i+r}$, defined previously,  which can be related to operators
$A \equiv (A_1^-,A_2^-, \cdots ,A_r^-, A_1^+, A_2^+, \cdots,
A_r^+)$ as follows
\begin{equation}
X = U \Omega^{-1} A
\end{equation}
where the matrix $\Omega$ (assumed to be invertible) is defined as
\begin{displaymath}
\Omega = \left( \begin{array} {ccc} u & v \\ v^* & u^*\\
\end{array} \right)
\end{displaymath}
and the matrix $U$ is given above. Using the transformation (69)
one can verify easily the following expressions of dispersion and
covariance matrices $\sigma (X)$ and $C(X)$
\begin{equation}
\sigma (X) = (U \Omega^{-1})\sigma (A) ({U\Omega^{-1}})^T
{\hskip 0.5cm}, {\hskip 0.5cm}
 C(X) = (U \Omega^{-1})C(A) ({U\Omega^{-1}})^T
\end{equation}
in terms of the dispersion and covariance of the $A$'s
operators. According the eigenvalues equations (68), the matrix
elements of $\sigma (A)$ and $C(A)$  are related by relations
similar to ones given by (60), (61) and (62). Then, one has  $\det
\sigma (A) = \det C(A)$ which implies  $\det \sigma (X) = \det
C(X)$.  Finally, we conclude that the Klauder-Perelomov  coherent
states, arising from the Bargmann realizations of bosonic and
fermionic $A_r$ statistics, minimize the Robertson-Shr\"odinger
uncertainty relation and they are, in this respect, intelligent.
\section{Conclusion}
This paper was devoted to the generalized $A_r$ statistics. We
have studied the associated Fock representations. We have obtained
the Fock spaces associated to bosonic ($s=1$) and fermionic
($s=-1$) $A_r$ statistics. In the limit $k \longrightarrow \infty$
($k$ index labeling the irreducible Fock representations), bosonic
as well as fermionic $A_r$ statistics reduce to the standard Bose
statistics. The $A_r$ statistics system becomes a collection of
ordinary bosons and the Jacobson generators coincide with creation
and annihilation operators of conventional degrees of freedom. We
have developed the Bargmann realizations of the Fock spaces and
determined the differential actions of the Jacobson generators. We
have shown that the so-called Klauder-Perelomov and Gazeau-Klauder
coherent states emerge, in a natural way, in these realizations.
The measures, by means of which we define the inner product of two
analytical functions for each considered realization, are
computed. They turn out to be the measures with respect which the
coherent states constitute over-complete sets.
We point out that the existence of two distinct Bargmann representations,
studied in sections 3 and 4, arises from choosing either the creation
 or the annihilation operator having a simple form similar to the ordinary
 Bose case; indeed in the latter the two coincide and there is only one Bargmann
 realisation, but they are necessarily distinct in the case under discussion.
  We shown also that
all obtained coherent states are intelligent. In other words, the states
give the minimum of the Robertson-Schr\"odinger uncertainty
relation. As first continuation, it would be interesting to study
a complete classification of intelligent states associated with
$A_r$ statistics. Furthermore, the results and tools presented in
this article can be extended to quantum statistics associated with
other classical Lie algebras and super-algebras. Finally, we
believe that the generalized $A_r$ statistics can be applied in
the study of quantum Hall effect in higher dimension spaces
[34,35]. We hope to report on this subject in a forthcoming work.
{\vskip 0.5cm}
\noindent {\bf Acknowledgements}\\
Thanks are due to the referees for pertinent and constructives remarks.

\end{document}